\documentclass[aps,prl,twocolumn,groupedaddress,amssymb]{revtex4-2}
\makeatletter
\def\set@curr@file#1{%
  \begingroup
    \escapechar\m@ne
    \xdef\@curr@file{\expandafter\string\csname #1\endcsname}%
  \endgroup
}
\def\quote@name#1{"\quote@@name#1\@gobble""}
\def\quote@@name#1"{#1\quote@@name}
\def\unquote@name#1{\quote@@name#1\@gobble"}
\makeatother
\usepackage{graphicx}
\usepackage{dcolumn}
\usepackage{bm}
\usepackage{amsmath}
\usepackage{xcolor}
\usepackage{epstopdf}
\usepackage{hyperref}
\usepackage{mwe} 
\usepackage{ulem}

\usepackage[utf8]{inputenc}

\newcommand{\probP}{\text{I\kern-0.15em P}}
\begin{document}

\preprint{APS/123-QED}

\title{Simple spatial processes can generate heterogeneous contact distributions in face-to-face interactions}

\author{Juliette Gambaudo}
 \email{juliette.gambaudo@cpt.univ-mrs.fr}
\author{Mathieu Génois}%

\affiliation{Aix Marseille Univ, Université de Toulon, CNRS, CPT, Marseille, France}%

\date{\today}

\begin{abstract}
Face-to-face interactions reveal recurring patterns, suggesting the possibility of shared underlying mechanisms. More specifically, inter-contact durations, contact durations and number of contacts per edge share similar heavy-tail distributions in many empirical settings. 
A common intuition is that face-to-face interactions may be influenced by spatial constraints, and that the observed complex behaviors could arise from such physical limitations. Our models explore the impact of this constraint by simulating pedestrian dynamics, and studying the generated temporal network of contacts. Previous work showed that the inter-contact duration distribution is recovered with a pedestrian dynamic as simple as the two dimensional random walk, but this approach doesn't allow to recover the distribution of the number of times a pair of individuals has been in contact. 
One assumption is that the number of contact between individual arises from the social relationship between them, in other words a memory of past interactions. However, we here present models that are based on solely spatial rules, by adding simple targeting mechanisms to the two-dimensional random walk. We show that these models allow to recover a broad distribution of the number of contacts, revealing the importance of two ingredients: localized phases and controlled population mixing. This suggests that the observed heterogeneity in the contact numbers within the data does not necessarily emerge from underlying social relationships between individuals, since an equivalent distribution may be reproduced using a purely spatially based model, without the need for memory mechanisms. 
\end{abstract}

\maketitle


\section{\label{sec:intro} Introduction}
Temporal networks collected in various settings exhibit heterogeneous behaviors \cite{holme2012temporal,karsaiBurstyHumanDynamics2018}. These temporal patterns are characterized by periods of high activity interspersed with long intervals of inactivity, a feature that deviates from Poissonian assumptions and calls for specific modeling efforts. In response, many network models have been developed aiming at reproducing these non-trivial temporal patterns \cite{barabasiOriginBurstsHeavy2005a, reisGenerativeModelsSimultaneously2020a, hiraokaModelingTemporalNetworks2020a, vazquezModelingBurstsHeavy2006a,joEmergenceBurstsCommunities2011, perraActivityDrivenModeling2012, bagrowNaturalEmergenceClusters2013, barratModelingTemporalNetworks2013}. Some of these models incorporate memory mechanisms, enabling them to account for the non-Markovian nature of real-world interactions \cite{vestergaard2014memory, karsaiUniversalFeaturesCorrelated2012, stehleDynamicalBurstyInteractions2010}.

\begin{figure}[ht!]
	\centering
	\includegraphics[width=\linewidth]{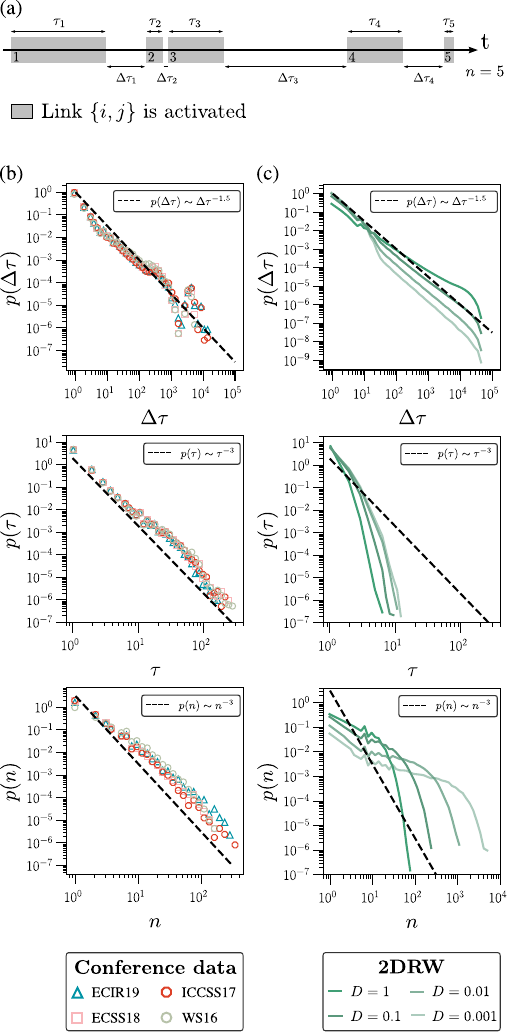}
	\caption{\textbf{a: Definition of the temporal observables.} Looking at the timeline of activation of the link between two nodes, one can define three observables: the contact durations $\tau$, the inter-contact durations $\Delta \tau$ and the number of contacts $n$. \textbf{b: Conference datasets.} In the conference dataset \cite{genoisCombiningSensorsSurveys2023}, those three observables are power-law distributed. \textbf{c: Two dimensional random walk.} Distributions obtained simulating $1000$ ghost particles undergoing a two dimensional random walk as defined in \cite{masoumi2025simplecrowddynamicsgenerate}. Only the inter-contact duration distribution is recovered.\cite{masoumi2025simplecrowddynamicsgenerate}}
	\label{confrontation_part}
\end{figure}

Among temporal networks, face-to-face interaction networks form a particularly interesting subset as they display universal heterogeneous features that have been consistently observed across multiple empirical studies \cite{cattuto2010dynamics, genoisCombiningSensorsSurveys2023}. Recent research focused on understanding the impact of heterogeneity in contact networks on the spreading of diseases \cite{masudaSmallIntereventTimes2020a, vazquezModelingBurstsHeavy2006a, miritelloDynamicalStrengthSocial2011, karsai2011small, panissonDynamicsHumanProximity2012, gauvinActivityClocksSpreading2013, holmeBirthDeathLinks2014, karsaiTimeVaryingNetworks2014}, motivating their study and understanding. Over the past decade, the SocioPatterns initiative has made significant contributions to this field by collecting high-resolution contact data in various social contexts \cite{sociopattern}. In these experiments, individuals are wearing proximity sensors and a contact is detected when two participants are facing each other in a close radius. These experiments allow for the reconstruction of temporal networks of contacts. On these networks, durations of contacts and between contacts, as well as the number of contacts between pairs of individuals, exhibit heterogeneous distributions, as seen in four different datasets collected during four different conferences \cite{genoisCombiningSensorsSurveys2023} (see Fig. \ref{confrontation_part}).

One promising modeling approach to capture the essence of face-to-face interaction networks is building networks of contacts from pedestrian dynamics. This idea emerges from the hypothesis that face-to-face interactions are strongly spatially constrained, a feature also explored in alternative frameworks such as spatiotemporal activity-driven networks \cite{simonSpatiotemporalActivityDrivenNetworks2025}, making it possible to capture their essential features through physical models of moving agents. The study of pedestrian dynamics is a very rich and established research area originating from physics and civil engineering \cite{hughesFLOWHUMANCROWDS2003, corbettaPhysicsHumanCrowds2023, cristianiMultiscaleModelingPedestrian2014, felicianiIntroductionCrowdManagement2021}. Initial physical models were inspired by Newtonian mechanics and included the notion of ``social forces", modeling individuals as particles subject to repulsive and attractive forces representing the will to move toward some target while keeping a distance within borders and other individuals \cite{helbingSocialForceModel1995}. Moussaïd et al. proposed a different approach, constructing their model on cognitive heuristics based on visual observations, allowing to reproduce crowd turbulence at high densities \cite{moussaidHowSimpleRules2011}.
 More recent developments modeled pedestrians with Langevin dynamics in a corridor \cite{corbettaFluctuationsMeanWalking2017}, in a curved path \cite{vleutenStochasticFluctuationsDiluted2024} adding pairwise avoidance \cite{corbettaPhysicsbasedModelingData2018} and finally in any geometric setting \cite{pouwDatadrivenPhysicsbasedModeling2024} capturing statistical properties of pedestrian dynamics.  Beyond motion modeling, moving agents have also been used as a framework to study dynamical processes, for instance in ecology \cite{orianaSimulatingPedestrianAvoidance2023} and epidemics \cite{PhysRevE.90.042813,Buscarino_2008}.

 Starnini et al. \cite{starnini2013modeling, starniniModelReproducesIndividual2016} bridged the gap between pedestrian dynamics and temporal networks, by analyzing the generated contact networks of particles following simple pedestrian dynamics. In their models, agents are assigned an attractiveness and their movement follow a two dimensional random walk \cite{rednerGuideFirstpassageProcesses2001} but their trajectories are biased by the attractiveness of the individuals they interact with. This model was sufficient to generate bursty contact sequence, and has been extended to study opinion dynamics \cite{starniniEmergenceMetapopulationsEcho2016a} or higher order face-to-face interactions \cite{galloHigherorderModelingFacetoface2024}. 
 
 Masoumi et al. \cite{masoumi2025simplecrowddynamicsgenerate} carried on this path with a more fundamental approach, simulating homogeneous particles, following three different dynamics: (i) the standard two-dimensional random walk (2DRW) \cite{rednerGuideFirstpassageProcesses2001}, (ii) Active Brownian particles (ABP) \cite{romanczuk2012active} which incorporates a short-term directional memory, and (iii) the Vicsek model that include short range interaction between individuals. Panel (c) of Fig. \ref{confrontation_part} displays the resulting distributions of contact durations, inter-contact durations and number of contacts obtained from the 2DRW model. Interestingly, they showed that even this minimal setup could reproduce the power-law distribution of inter-contact durations. This result allowed them, using the work of Rast \cite{rastContactStatisticsPopulations2022}, to link the exponent $-1.5$ of the distribution seen in the data to the one of the first passage time of the one dimensional random walk \cite{rednerGuideFirstpassageProcesses2001} with a semi-analytic argument. However, both 2DRW, ABP and Vicsek models failed to reproduce the heavy-tail distributions for contact durations and number of contacts, revealing the need of additional mechanisms beyond simple motion dynamics.

 In this paper, we propose additional simple targeting rules allowing to recover heterogeneous numbers of contact distributions. In the presented models, particles alternate between phases of free diffusion and phases of diffusion with a drift targeting a point in space. We investigate different target-choice mechanisms, allowing to understand the fundamental ingredients necessary for generating heterogeneous number of contacts. We show that having long localized phases while keeping a balanced population mixing allows to recover heavy tail distributions. This suggest that the heterogeneity of the number of contacts among pairs of individuals observed in the data does not necessarily emerge from a memory of past interactions, as simple targeting mechanisms allow to generate it. 

\section{\label{sec:method}Methods}
\subsection{\label{subsec:data}Data}
We use the data collected by Génois et al. \cite{genoisCombiningSensorsSurveys2023} as a comparison baseline. These datasets were collected in four different conferences: the 3rd GESIS Computational Social Science Winter Symposium of 2016 (WS16), the International Conference of Computational Social Sciences of 2017 (ICCSS17) and the Eurosymposium on Computational Social Science in 2018 and 2019 (ECIR18 and ECIR19). Contact were collected through the Sociopattern platform \cite{sociopattern}, where participants' nametags are equipped with RFID sensors and antennas are distributed throughout the venue to capture contact data. Each sensor detects others within $\simeq 1.5$ meters, but only when individuals face each other, as the human body blocks the signal outside of the front half-sphere. A contact is defined by this proximity and orientation, recorded every 20 seconds. In these datasets, the number of participants for which contact were detected varies between $138$ and $274$ \cite{genoisCombiningSensorsSurveys2023}. Looking at the timeline of activation of the link between two nodes, one can define three observables: the contact durations $\tau$, the inter-contact durations $\Delta \tau$, the duration between two consecutive contacts, and the number of contacts $n$, defined per pairs of individuals (see panel (a) of Fig. \ref{confrontation_part}). These temporal distributions observed in these datasets are plotted in Fig. \ref{confrontation_part} (b). The temporal unit on the horizontal axis for contact and inter-contact durations is in number of time-steps. They exhibit similar heavy tail distributions over several orders of magnitude, hinting towards the existence of an universal behavior.

\subsection{\label{subsec:model}Models}
\subsubsection{\label{subsubsec:peddyn}Turning pedestrian dynamics into temporal networks}
\begin{figure}[p]
	\centering
	\includegraphics[width=\linewidth]{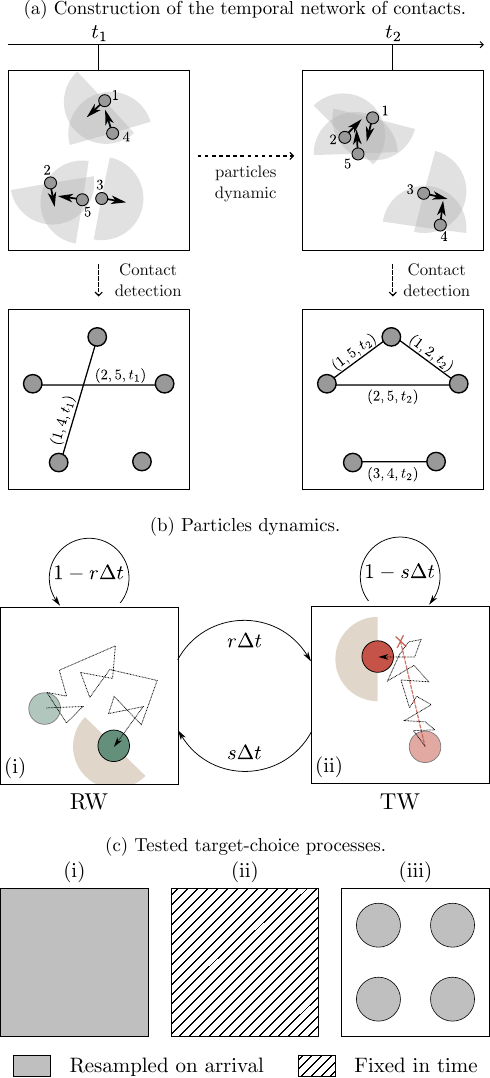} 
	\caption{\textbf{Description of the models.} \textbf{a:} Particles are moving following simple pedestrian dynamics and at each time-steps a contact is defined when two particles are facing each other within a certain radius. \textbf{b:} A particle alternates between (i) a Random Walk (RW), where the particle is freely diffusing and (ii) a Targeting Walk (TW), where the particle's motion is biased with a drift towards a target. The transition rates between the two walks are $r$ and $s$. \textbf{c:} When a particle is performing a Targeting Walk, three target position choice processes are tested: (i) Resampled on arrival destination. (ii)  Fixed in time destination and (iii) Constrained resampled on arrival destination.}
	\label{fig:methods}
\end{figure}

We consider $N$ particles performing a two-dimensional walk in a continuous space with discrete time. We define $P=\{1, \dots, N\}$ the set of particles. The time increment $\Delta t$  is set to $1$ and the total amount of time steps is $n_t$, defining the set of time steps $T = \{\Delta t, \dots, n_t\Delta t\}$. The space is a squared box of size $L$. A contact between two particles is defined when they are facing each other within a certain radius in their front half-sphere, to match the data collection setting. An example of contact detection is schematized in panel (a) of Fig. \ref{fig:methods} where at time-step $t_1$ particles $2$ and $5$ are within each other's front-half sphere, defining a contact, but particles $3$ and $5$, while being within each other's detection radius, are not facing each other and so no contact is detected between them. Based on this definition of contact, the temporal network of contacts between particles is built, where nodes correspond to particles and links are the detected contacts. See \cite{masoumi2025simplecrowddynamicsgenerate} for further details of the base model.

\subsubsection{\label{subsubsec:part_dyn}Particles dynamics}

In our model, each particle alternates between two walks: a Random Walk (RW), and a Targeting Walk (TW). At each time-step $t \in T$, we define two sub-sets of $P$, $RW(t)$ the set of particles performing the Random Walk at time $t$ and $TW(t)$ the set of particles performing the Targeting Walk at time $t$. The populations forming the two sets $RW(t)$ and $TW(t)$ evolve through time, with a constant rule that $P$ is the disjoint union of $RW(t)$ and $TW(t)$ at each time-step:
\begin{equation} \label{eq:states_cst}
	\forall t \in T, ~
	\begin{cases}
      	RW(t) \cup TW(t) = P\\
	    RW(t) \cap TW(t) = \emptyset
    \end{cases}\,.
\end{equation}

 The alternation between the two states proceeds as follows. A particles switches from a random to a targeting walk with rate $r$ and in reverse with rate $s$:

 $\forall i \in P, ~ \forall t \in T,$
\begin{equation} \label{eq:states_proba}
	\begin{cases}
      \probP\left(i \in TW(t+ \Delta t)|i \in RW(t)\right) = r \Delta t\\
	  \probP\left(i \in RW(t+ \Delta t)| i \in TW(t)\right) = s \Delta t
    \end{cases}\,.
\end{equation}
where $\probP$ is the conditional probability for a particle to switch walk between two consecutive times-steps. The described process is schematized in panel (b) of Fig. \ref{fig:methods}. Depending on the walk, the particle's position is updated differently.

\paragraph{\label{para:2DRW} The Random Walk (RW).}
At each time step, each particle executing a Random Walk choses uniformly a new direction $\Theta$ and draws a step length $\Delta r$ from a semi-gaussian centered around zero and of standard deviation $D \times \Delta t$. The position of particle $i \in  RW(t)$, $\mathbf{r}_i(t)$ is updated as follow:

$\forall i \in RW(t),$
\begin{equation}\label{eq:2DRW}
	 \mathbf{r}_i(t+\Delta t) = \mathbf{r}_i(t) + \Delta r_i(t) \binom{\cos \Theta_{i} (t)}{\sin \Theta_{i} (t)}
\end{equation}

\paragraph{\label{para:Targeting} The Targeting Walk (TW).}
In this state, a particle's walk is biased by adding a drift $\mathbf{u}(\mathbf{r}_i(t))$ always pointing towards a same point, the target. We test different processes to define the targets' positions $\{r_{T_i}\}_{i \in  TW(t) }$, their descriptions is detailed in the next section. Additionally to this drift, the particle's movement is also diffusing following the same rules are in the case of the random walk.

$\forall i \in TW(t), $
\begin{equation}\label{eq:2DRW_drift}
	\mathbf{r}_i(t+\Delta t) = \mathbf{r}_i(t) + \mathbf{u}(\mathbf{r}_i(t)) + \Delta r_i(t) \binom{\cos \Theta_{i} (t)}{\sin \Theta_{i} (t)}
\end{equation}
Where $\mathbf{u}(\mathbf{r}_i(t))$, the normalized vector pointing towards the target, is defined as follows:
\begin{equation}\label{eq:drift}
	\mathbf{u}(\mathbf{r}_i(t)) =\dfrac{\mathbf{r}_{T_i} - \mathbf{r}_i(t)}{||\mathbf{r}_{T_i} - \mathbf{r}_i(t)||}
\end{equation}
$\mathbf{r}_{T_i}$ being the position of the target of particle $i$.

\subsubsection{\label{subsubsec:choice_target}Choice of the target}
The target is a position in the space that is assigned to a particle. When the particle is executing a Targeting Walk, it adds a drift to its diffusion pointing towards this target (Eq. \ref{eq:2DRW_drift}). Here, we detail three different mechanisms to assign positions to the targets.

\paragraph{\label{para:full_spatial}  (i) Resampled on arrival destination.} 
In this case, the position of the target is drawn randomly in the whole box on arrival into the $TW$ state, meaning that each time a particle switches from a Random Walk to a Targeting walk, the target is reset to a new random position. Therefore, in this case, particles change target through time, with periods of free diffusion between two consecutive targets.

\paragraph{\label{para:home}  (ii) Fixed in time destination.}
In this case, the position of the target is drawn randomly in the whole box at the beginning of the simulation. It is also taken to be the initial position of the particle in the simulation. The target position remains identical throughout the full duration of the simulation. Therefore, particles alternate between localized periods around the same point and free diffusion periods. 

\paragraph{\label{para:poi} (iii) Constrained resampled on arrival destination.}
In this case, areas of the space are defined as ``areas of interest" and initial target positions are drawn from these areas. Each time a particle switches from the random walk state to the targeting state, the target is reset to a new random position from one of these areas. Two consecutive choice of target positions can be drawn in different areas. This case derives from the \textit{Resampled on arrival} with an additional constraint on the potential position of target. This intends to be closer to observed behavior in face-to-face interactions, such as the buffet areas in a conference. 

Panel (c) of Fig. \ref{fig:methods} summarizes the three target-choice mechanisms presented in this paper.

\section{\label{sec:results}Results}
\subsection{\label{subsec:main_results} Number of contact distributions for different targeting mechanisms in different regimes}
\begin{figure*}[p]
	\centering
	\includegraphics[width=\linewidth]{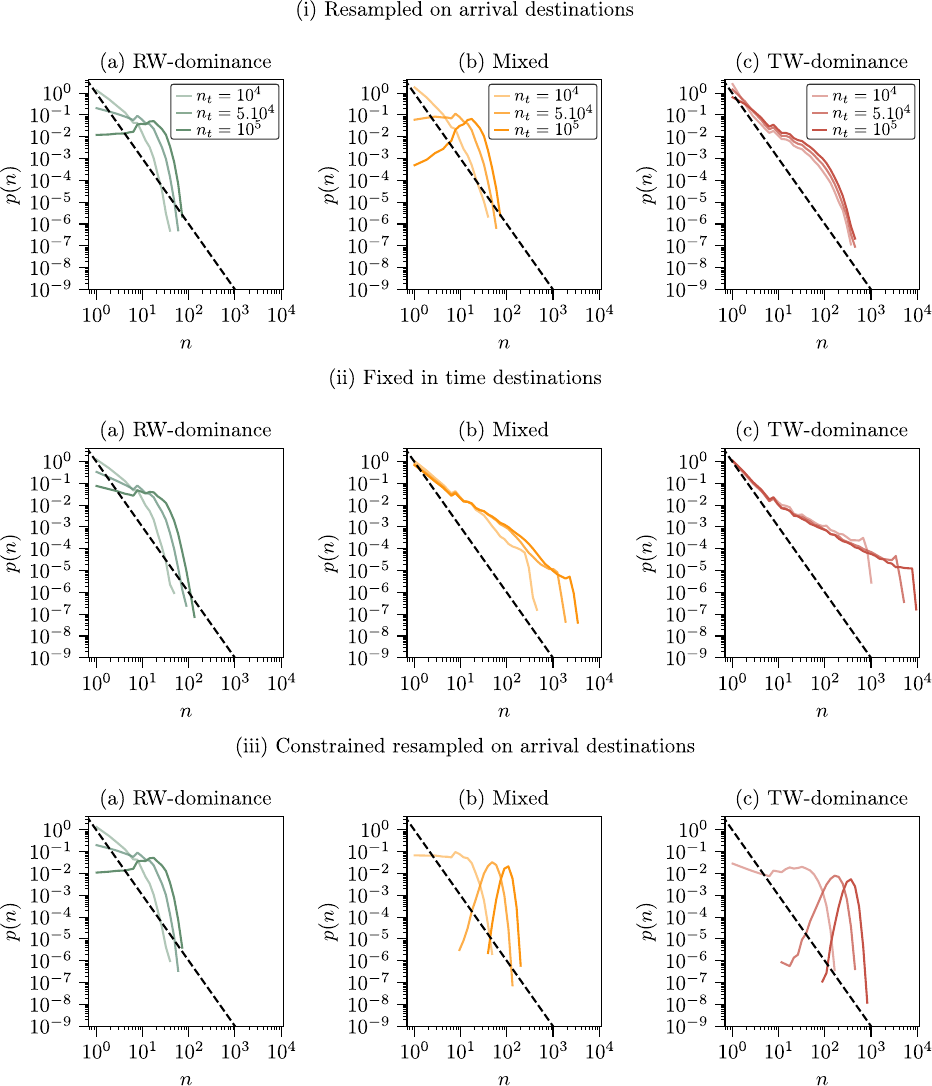} 
	\caption{\textbf{Stability of the number of contact distributions.} Each row presents the result for a different targeting mechanism. The first row corresponds to \textit{resampled on arrival destinations}, the second to \textit{fixed in time destinations} and the last to \textit{constrained resampled on arrival destinations}. Each column correspond to different values of the rates $r$ and $s$. The first column corresponds to the $RW$-dominance case ($r=0.001$ and $s=0.1$), the second column corresponds to the mixed case ($r=s=0.01$) and the last column to the $TW$-dominance case ($r=0.1$ and $s=0.001$). In each cases is plotted the distribution for different values of $n_t$, the total amount of time-steps. The dashed ine is the slope $p(n) \sim n^{-3}$ present in Fig. \ref{confrontation_part}.}
	\label{fig:res_main}
\end{figure*}

We run simulations for $N=1000$ ghost particles in a box of size $L=100$ with reflecting boundary conditions during $t=10^5$ time steps. The detection radius of the particles is set to $r_D=1.5$, and the diffusion coefficient is set to $D=1$. In Supplementary Materials, we show that the choice of $N$ and $D$ only impacts the statistics and not the shape of the distribution. For \textit{constrained resampled on arrival} target-choice mechanism, we choose the case of four areas of interest of radius $15$ distributed as schematized in panel (c. iii) of Fig. \ref{fig:methods}. For each targeting process (corresponding to each row of Fig. \ref{fig:res_main}), three extreme cases are tested (and correspond to each column of Fig. \ref{fig:res_main}):
\paragraph{(a) The $RW$-dominance case.} Here, $r=10^{-3}$ and $s=10^{-1}$, meaning that particles are freely diffusing most of the time.

\paragraph{(b) The mixed case.} Here  $r=s=10^{-2}$, meaning that particles spend as much time performing a Random Walk as a Targeting Walk.

\paragraph{(c) The $TW$-dominance case.} Here $r=10^{-1}$ and $s=10^{-3}$, meaning that particles are spending most of their time performing Targeting Walks than Random Walks.

On Fig. \ref{fig:res_main}, we show the evolution of the distribution of number of contacts per pair of particles in each case by computing the distribution obtained for different simulation durations. For the first two targeting mechanisms (\textit{Resampled on arrival} and \textit{Fixed in time} destinations, in the first two rows) the distribution broadens when the targeting walk becomes dominant. With \textit{fixed in time} target-choice, the tail is heavier than when the target is \textit{resampled on arrival}. This is due to pairs of particles having their fixed targets close to each other interacting a very large amount of time. This effect is minimized for \textit{resampled on arrival} targets because particles change target after switching state. With the third targeting mechanism, when destinations are \textit{constrained and resampled on arrival} (row (iii) of Fig. \ref{fig:res_main}), the distribution doesn't get broader, but shifts its mean toward higher number of contacts, translating the absence of pairs of particles with low number of contacts. In this case, having areas of interest increases the probability for two particles to meet, since resampled targets can be chosen from any of the areas, resulting in a high mixing of the population.

Concerning the stability of the distributions, results show that the impact  of the duration of the simulation is different depending of the targeting mechanism and the regime. In the $RW$-dominated regime (first column) increasing the duration of the simulations broadens the distributions, at the cost of loosing occurrences of small values of $n$. This is due to the fact that pairs of particles have more opportunities to meet a higher amount of times for longer simulations. This effect is amplified when targets are \textit{constrained and resampled on arrival} (row (iii)), where increasing the duration of the simulations narrows and shifts the mean of the distribution eventually resulting in a narrower distribution centered around a high value of $n$. This demonstrates that, in such cases, particles exhibit excessive exploration, leading to a rarefaction of particle pairs that have interacted for only small number of times. For \textit{fixed in time destinations}, in the mixed and $TW$-dominated regimes, increasing the amount of time-steps broadens the tail of the distribution, without loosing occurrences of small number of contacts.  

\begin{figure*}[ht!]
	\centering
	\includegraphics[width=\linewidth]{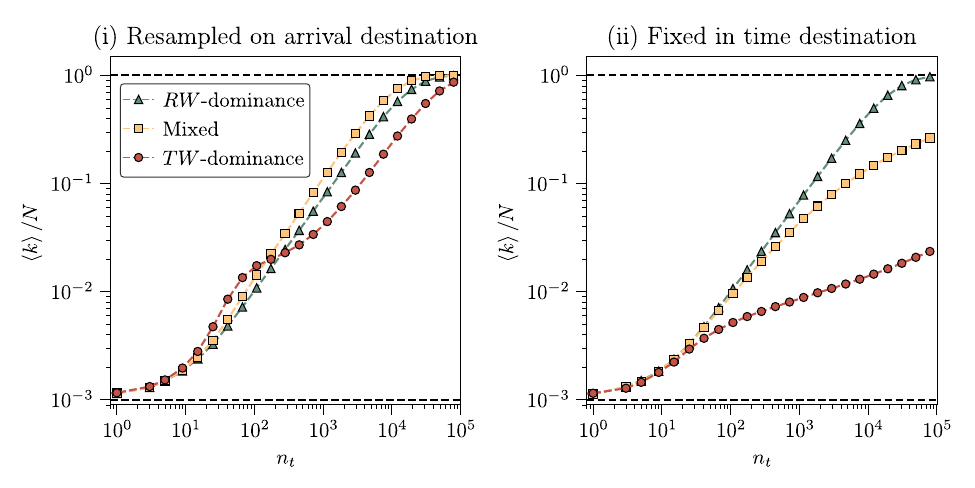} 
	\caption{\textbf{Evolution of the average degree in the aggregated network with two homogeneous target-choice mechanism.} Each panel correspond to different targeting mechanisms: (i) resampled on arrival destinations and (ii) fixed in time destination (see panel (c) Fig. \ref{fig:methods}). Both panels show results for three different regime: the $RW$-dominance case ($r=0.001$ and $s=0.1$), the $TW$-dominance case ($r=0.1$ and $s=0.001$) and the mixed case ($r=s=0.01$). The two horizontal dashed lines show the domain of evolution of $\left<k\right>/N$, i.e. between $10^{-3}$ (corresponding to $\left<k\right> = 1$) and $1$ (corresponding to $\left<k\right> = N$). Results were obtained by averaging $10$ simulations of $n_t=10^5$ time-steps with $N=1000$ ghost particles. Errors were computed and are too low to appear on the plot.}
	\label{fig:res_full_spatial_and_home_loglog_evol_kmean}
\end{figure*}

\subsection{\label{subsec:kmean}Evolution of the mean degree of the aggregated network}

For the different cases tested, studying the structure of the aggregated network gives more insight into topological effects of each mechanism. Here, we examine the temporal evolution of the mean degree in the aggregated network (Fig. \ref{fig:res_full_spatial_and_home_loglog_evol_kmean}). Both panels of Fig. \ref{fig:res_full_spatial_and_home_loglog_evol_kmean} show the evolution of the average degree for the three different regimes, in the left panel particle performing the $TW$ target \textit{resampled on arrival destinations} (case (i) of panel (c) of Fig. \ref{fig:methods}), while in the right panel, they target \textit{fixed in time destinations} (case (ii) of panel (c) of Fig. \ref{fig:methods}). We do not show studies of the aggregated network for the third configuration (\textit{constrained resampled on arrival destinations}), as in this case the networks quickly becomes fully connected. 

The first observation that can be made on these results is wether the average degree in the aggregated network reaches the saturation value corresponding to a fully connected network (horizontal dashed line $\left<k\right>/N = 1$). If \textit{resampled on arrival} targets all lead to curves approaching this threshold, indicating that all particles interacted with most of the other particles, this is not the case when the target is \textit{fixed in time}. Actually, the more particle spend time performing a $TW$ rather than a $RW$, the less likely they are to be in contact with all other particles during the simulation. Indeed, in this state the explorable space is constrained around the target, and since the target position is fixed in time, particles can only explore outside of the vicinity of this target when they switch to the $RW$.

The second observation that can be made relates to the rate at which the average degree increases. Interestingly, with the \textit{resampled on arrival} target-choice mechanism, the average degree presents two domain. For $0\leq n_t \leq 10^2$, $TW$-dominance case presents a faster growth of the average degree, while the $RW$-dominance produces the slowest growth, and the mixed case generates an intermediary growth. This faster growth is due to particle entering the $TW$ crossing the space to reach their target, and therefore crossing the path of other particles. This doesn't happen in the right panel, because the \textit{fixed in time destination} is chosen to be the initial position of the particle. For $ n_t \geq 10^2$, the growth of the average degree in the $TW$-dominance case slows down. Interestingly, this is not the case for the mixed case. This shows the impact of the choice of a new target being on arrival rather than fixed through time. Indeed, in this case compared to the $TW$-dominance one, particles enter a larger amount of time into the $TW$ state, since they had more occasions to switch to the $RW$ and back, leading on each occurrence to a new choice of target and eventually to a higher mixing of the population.

\section{\label{sec:discussion}Conclusion}
Presented results indicate two key ingredients that allow these model to generate heterogeneous number of contacts between pairs of individuals: long localized phases and controlled population mixing.
\subsection{\label{subsec:discussion1}Long localized phases can generate broad number of contacts distributions.}

The presented models show the impact of adding localized phases to random walking particles on the distribution of the number of contacts between pairs of particles. Three targeting processes were tested. We focus here on the $TW$-dominated regime when targets are \textit{resampled on arrival} and on the mixed and $TW$-dominated regimes when targets are \textit{fixed in time}, where a stable heavy tail is recovered. In both cases, particles alternate between freely diffusing phases and constrained phases where the targets' positions are homogeneously distributed in the space. Heavy tail distributions of the number of contacts emerge when particles spend long periods in the localized state. The difference between the case where the targeting is \textit{resampled on arrival} or \textit{fixed in time} remains in the persistence of the target's position through time. When targets are \textit{resampled on arrival}, the choice of the target is randomly selected for each time entering the $TW$ state, which is not the case when the targets positions are \textit{fixed in time}. If in the latter, the fixed target can be interpreted as a form of memory, since the link timeline between pairs of particles will depend strongly on the proximity of their targets, it is important to point out that this is not the case with the first targeting mechanism. This reveals the key role that the localized phases have on the generation of the distribution, that is more fundamental than a memory mechanism.

\subsection{\label{subsec:discussion2}The importance of the position of the localized phases to maintain a balanced population mixing.}
When targets are \textit{resampled on arrival} or \textit{fixed in time}, a stable heavy tail is recovered. In both of these cases, the targets' positions are homogeneously distributed in the space. This is not the case when the target choice is \textit{constrained and resampled on arrival}, where the targets' positions are selected from defined perimeters within the box in which  particles are evolving. Results show how this additional constraint prevents from recovering a heavy tail distribution, but rather produce a system where all particles meet all other particles a high amount of time. Indeed, in this case, since all particles choose their targets within the same areas, the population mixing increases. If the first two cases showed how long localized phases could generate heavy tail distribution of the number of contacts, this last case showcases the importance of these localized phases to be homogeneously distributed in the accessible space.

This conclusion is also confirmed by studying particles behaviors in the mixed regime when the target choice is \textit{resampled on arrival}, compared to the behavior in the same regime with the \textit{fixed in time} target-choice mechanism. In the latter, the average degree in the aggregated network (see right panel of Fig. \ref{fig:res_full_spatial_and_home_loglog_evol_kmean}) has an intermediary growth between the two extreme cases. This is not the case for the mixed regime with the \textit{resampled on arrival} target-choice mechanism, where the average degree of the aggregated network saturates faster than in the $RW$-dominated regime (see left panel of Fig. \ref{fig:res_full_spatial_and_home_loglog_evol_kmean}). As discussed in results section, this shows how choosing a new target on arrival in the $TW$ state increases the population mixing. Looking at the resulted distribution of number of contacts per edges in  this case, the effect of the mixing becomes visible, with a distribution converging faster toward a high mean narrow distribution.

Finally, the position of the localized phases impacts the slope of the distribution. Indeed, depending on if the targets are \textit{resampled on arrival} or \textit{fixed in time}, the recovered heavy tail doesn't have the same slope. This reveals how the slope is determined by the population mixing induced by the change of target through time.
\section{\label{sec:conclusion}Discussion}

We conducted contact studies of pedestrian dynamics governed by simple spatial rules, aiming to reduce the complex problem of heterogeneous number of contacts into a minimal model. To compare model results and empirical data, we performed qualitative comparisons. However, attempt at reproducing the same distributions would call for quantitative analysis and comparison of the distributions.

We found that, despite the lack of realism of these models, they allow to generate similar behavior as observed in face-to-face data. Indeed, in the presented models, particles are homogeneous and memory-less. Particles trajectories are close to the one of a two dimensional random walk and do not follow realistic trajectories as studied in the crowd dynamic literature \cite{corbettaPhysicsHumanCrowds2023}. Yet, they allow to reproduce complex behaviors, such as the observed power-law distribution of inter-contact durations \cite{masoumi2025simplecrowddynamicsgenerate}, and an heterogeneous distribution of number of contact between pairs of individuals. Retrieving the latter result with memory-less particles challenges initial intuitions that the number of time two individuals are in contact relies on the long-lasting relationship between these two individuals. We show that, on the contrary, simple spatial rules are enough to retrieve the same behavior, indicating that there is insufficient evidence to attribute the number of contact observed in the face-to-face data to individuals' social relationships.

This work shows that two ingredients are necessary to generate heterogeneous number of contacts between pairs of particles with only spatial rules: long localized phases and controlled population mixing. We highlighted the first ingredient by testing models where particles alternate between free diffusion and targeting walk. We showed that these phases allow to generate stable heavy-tail distributions. Our models do not allow to recover the same slope as the one observed in the data, but our work shows that the slope of the tail depends on the targeting mechanism, giving additional insight on how to interpret the empirical distribution. The importance of the second ingredient, a limited population mixing, was showcased through the study of various targeting mechanisms, showing that those that induced higher mixing of the population were not recovering stable heavy-tailed distribution of $n$. In the context of conferences, these findings may be interpreted as the observed heterogeneity in empirical data partly arise from participants alternating between static and diffusive behavioral phases, where static phases would include activities such as engaging in conversations during coffee breaks, interacting with a presenter at a poster session, or simply sitting for lunch, while diffusive phases would involve wandering among posters, transitioning between presentation rooms, or seeking spaces for discussion.

This work highlights the pertinence of using simple pedestrian dynamics to model temporal networks of contact. Indeed, their inherent spatial complexity, that they share with face-to-face interactions, allows to recover complex behaviors with few additional simple rules. Further studies could therefore make use of pedestrian dynamics to explore other observed phenomena on temporal network of contacts, such as the heavy tail behavior of the contact duration distribution (see third row of panel (b) of Fig. \ref{confrontation_part}) or temporal correlations of events in the temporal network. Extended research could also investigate how the spatial configuration of the environment, such as room shape, presence of an obstacle or global movement from one room to another through a door influences the temporal network's dynamic.

\section*{Acknowledgements}
\subsection*{Author contributions}
M.G. proposed the core idea of the project. J.G. and M.G. contributed to the scientific discussions of the work. J.G. wrote code, performed numerical simulations and analyzed numerical and empirical data. J.G. and M.G. wrote and reviewed the final manuscript.

\subsection*{Funding}
J.G. and M.G. are partially supported by the Agence Nationale de la Recherche (ANR) project DATAREDUX (ANR-19-CE46-0008) and by the ANR France 2030 project Data2Laws (24-EXEN-0002).

\subsection*{Competing interests}
The authors declare no competing interests.

\bibliography{mybib}

\newpage
\onecolumngrid

\end{document}